\documentclass[aps,prl,twocolumn,showpacs,
superscriptaddress,amsmath,amssymb]{revtex4-1}
\usepackage{graphicx}
\usepackage{epsfig}
\usepackage{dcolumn}
\usepackage{bm}
\usepackage{pslatex}

\begin{document}
\title{Fatigue in disordered media}

\author{Cl\'audio L. N. Oliveira}%
\email{lucas@fisica.ufc.br}%
\affiliation{Departamento de F\'isica, Universidade Federal do
  Cear\'a, 60451-970 Fortaleza, Cear\'a, Brazil}

\author{Andr\'e P. Vieira}%
\email{apvieira@if.usp.br}%
\affiliation{Instituto de F\'isica, Universidade de S\~ao Paulo,
  05314-970, S\~ao Paulo, Brazil}

\author{Hans J. Herrmann}%
\email{hans@ifb.baug.ethz.ch}%
\affiliation{Departamento de F\'isica, Universidade Federal do
  Cear\'a, 60451-970 Fortaleza, Cear\'a, Brazil}%
\affiliation{Computational Physics, IfB, ETH Z\"urich, H\"onggerberg,
  CH-8093 Z\"urich, Switzerland}

\author{Jos\'e S. Andrade Jr.}%
\email{soares@fisica.ufc.br}%
\affiliation{Departamento de F\'isica, Universidade Federal do
  Cear\'a, 60451-970 Fortaleza, Cear\'a, Brazil}


\begin{abstract}
  We obtain the Paris law of fatigue crack propagation in a disordered
  solid using a fuse network model where the accumulated damage in
  each resistor increases with time as a power law of the local
  current amplitude. When a resistor reaches its fatigue threshold, it
  burns irreversibly. Over time, this drives cracks to grow until the
  system is fractured in two parts. We study the relation between the
  macroscopic exponent of the crack growth rate -- entering the
  phenomenological Paris law -- and the microscopic
  damage-accumulation exponent, $\gamma$, under the influence of
  disorder. The way the jumps of the growing crack, $\Delta a$, and
  the waiting-time between successive breaks, $\Delta t$, depend on
  the type of material, via $\gamma$, are also investigated. We find
  that the averages of these quantities, $\left\langle\Delta
    a\right\rangle$ and $\left\langle\Delta
    t\right\rangle/\left\langle t_r\right\rangle$, scale as power laws
  of the crack length $a$, $\left\langle\Delta a\right\rangle \propto
  a^{\alpha}$ and $\left\langle\Delta t\right\rangle/\left\langle
    t_r\right\rangle\propto a^{-\beta}$, where $\left\langle
    t_r\right\rangle$ is the average rupture time. Strikingly, our
  results show, for small values of $\gamma$, a decrease in the
  exponent of the Paris law in comparison with the homogeneous case,
  leading to an increase in the lifetime of breaking materials.  For
  the particular case of $\gamma=0$, when fatigue is exclusively ruled
  by disorder, an analytical treatment confirms the results obtained
  by simulation.
\end{abstract}


\maketitle

Fatigue failure is relevant in most everyday-life situations.  For
instance, asphalt pavement has its lifetime decreased by cyclic
traffic loading, and aircraft fuselage has its lifespan determined by
the number of pressurization cycles \cite{Kanninen1985}. Fatigue is
the largest cause of engineering failure, comprising an estimated 90\%
of all metallic failures \cite{Callister2007}. Furthermore, due to its
non-linear and critical behavior, it is mostly catastrophic and
insidious, occurring very suddenly and without warning if the crack is
not noticed, since it is brittle-like even in normally ductile
materials \cite{Picallo2010, Herrmann1988}. So far, theoretical
understanding of crack-growth phenomena is far from complete, and
fatigue-life prediction remains essentially empirical \cite{Kun2007}.

Generally speaking, failure in stressed materials is the result of
accumulated microscopic damage due to cyclic or constant external
loads. For cyclic (constant) load one speaks of fatigue (creep)
failure \cite{Alava2011, Kun2009}.  In a mesoscopic description, when
the accumulated microscopic damage reaches a local threshold, cracks
are initiated and grow through the material, leading to macroscopic
fracture. Under fatigue, local damage increases and cracks grow even
for external loads below the limit strength of the material. Crack
growth in this sub-critical regime, for an intermediate crack length
$a$, is usually described by the empirical Paris law \cite{Paris1963},
\begin{equation}
  \frac{da}{dt} = C(\Delta K)^m \propto a^{m/2},
  \label{eq:parislaw}
\end{equation}
which relates the crack velocity $da/dt$ with the stress-intensity
amplitude $\Delta K\propto\Delta\sigma\sqrt{a}$, for a sample subject
to an external stress amplitude $\Delta\sigma$. The constants $C$ and
$m$ depend on the material and on the loading conditions
\cite{Kanninen1985}.

Due to the intrinsic characteristics of random microscopic defects in
fracturing materials \cite{Hansen2010, Lazarus2011, Altus2012,
  Manzato2012, Papanikolaou2011, Rosti2010, Niccolini2009,
  Zapperi2007}, statistical models such as the random spring model
\cite{Sahimi1996}, the random beam model \cite{Kun2007, Kun2008}, and
the random fuse model \cite{Herrmann1988, Hansen1991, Zapperi2009,
  Arcangelis1989} have been successfully applied to study fracture.
The role of disorder in these systems is to induce a transition from
brittle to ductile behavior \cite{Herrmann1988, Roux1988,
  Gilabert1987, Petri1994, Moreira2012}. As already mentioned, the
presence of fatigue can also lead to brittleness, although the
physical mechanism responsible for this behavior so far remains
unknown \cite{Callister2007}. Vieira {\it et al.}  \cite{Vieira2008}
have shown analytically and confirmed numerically the relation between
the Paris-law exponent $m$ and the microscopic damage-accumulation
parameter $\gamma$ in homogeneous materials. In the same study, first
attempts have been made to analyze the stability of the obtained
results towards the introduction of disorder. The aim of the present
work is to provide the full picture, using fuse-model simulations and
analytical arguments, of the role of disorder in fatigue crack-growth
phenomena, including the Paris as well as the catastrophic
(post-Paris) regime.

In our implementation, the random fuse model consists of resistors
(with the same conductance) located on the bonds of a tilted square
lattice, mimicking springs in the more complicated case of fracture
mechanics \cite{Herrmann1988, Gilabert1987}. In this scalar version of
elasticity, the current (voltage) in the electrical system is
analogous to the magnitude of the stress (strain) in the mechanical
system. In order to simulate heterogeneous microscopic defects
commonly found in real materials, we assign to each resistor a random
fatigue threshold $F^{\mathrm{th}}$, chosen from a uniform probability
distribution between $1-b$ and $1+b$, where $b$ gauges the strength of
disorder. Initially, the system has a notch, formed by removing two
contiguous bonds, placed symmetrically at the center of the system,
with periodic boundary conditions along the notch direction. A global
transverse current is imposed across the system from top to bottom and
the local currents are numerically calculated by applying Kirchhoff's
law to each node \cite{HSL} (nodes being the points where four
resistors meet). We then compute, for each resistor $i$, the fatigue
damage history up to time $t$ by using a power-law damage-accumulation
function,
\begin{equation}
  f_i(t) \propto \int_0^{t} \left[I_i(t^{\prime})\right]^{\gamma}dt^{\prime},
\end{equation}
where $I_i(t)$ is the current in resistor $i$ at time $t$, and the
stress-amplification exponent $\gamma$ is a phenomenological
microscopic parameter, dependent on the material. Below the fatigue
threshold, a resistor follows Ohm's law, but once fatigue damage
reaches the corresponding threshold, $f_i=F_i^{th}$, the resistor
burns irreversibly, like an electrical fuse. The system here is
operated in the quasi-static limit, i.e., the period of the current
cycle is much larger than the time required for the rearrangement of
the current distribution following the burning of a fuse, but much
smaller than the time between two consecutive burning events. This is
equivalent to assuming a well-defined current amplitude $\Delta I_i$
in each resistor during that time.  Therefore, the damage increment at
resistor $i$ between two consecutive burning events occurring at times
$t$ and $t+\Delta t$ is simply given by
\begin{equation}
  \Delta f_i(\Delta t) = A\Delta t\left[\Delta I_i\left(t\right)\right]^{\gamma},
\end{equation}
where $\Delta I_i\left(t\right)$ is the current amplitude in resistor
$i$ between events and the constant $A$ sets the time scale so that
the damage $f_i$ becomes dimensionless.

\begin{figure}[t]
  \begin{center}
    \includegraphics[width=0.8\columnwidth]{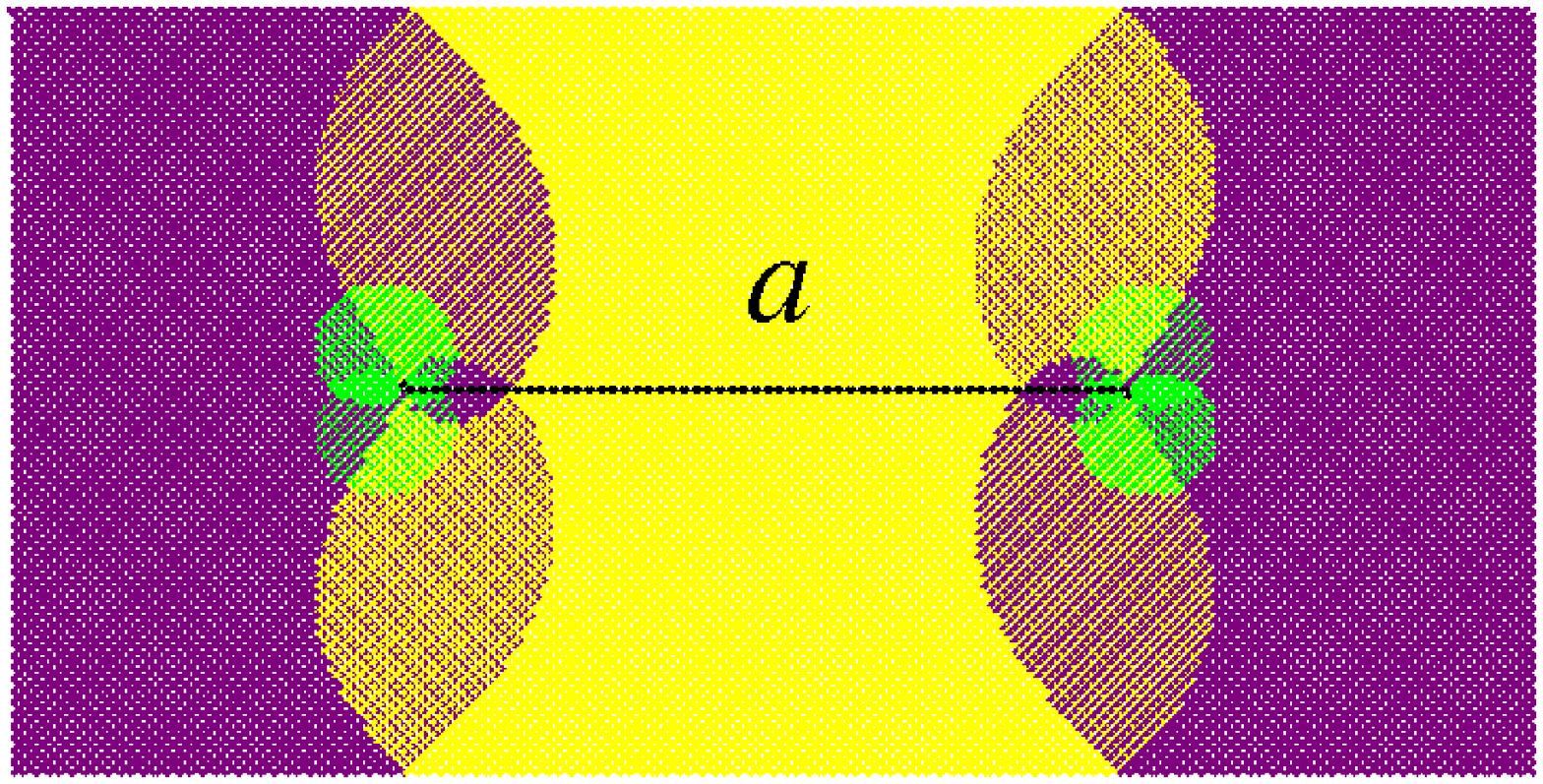}
    \includegraphics*[width=0.8\columnwidth]{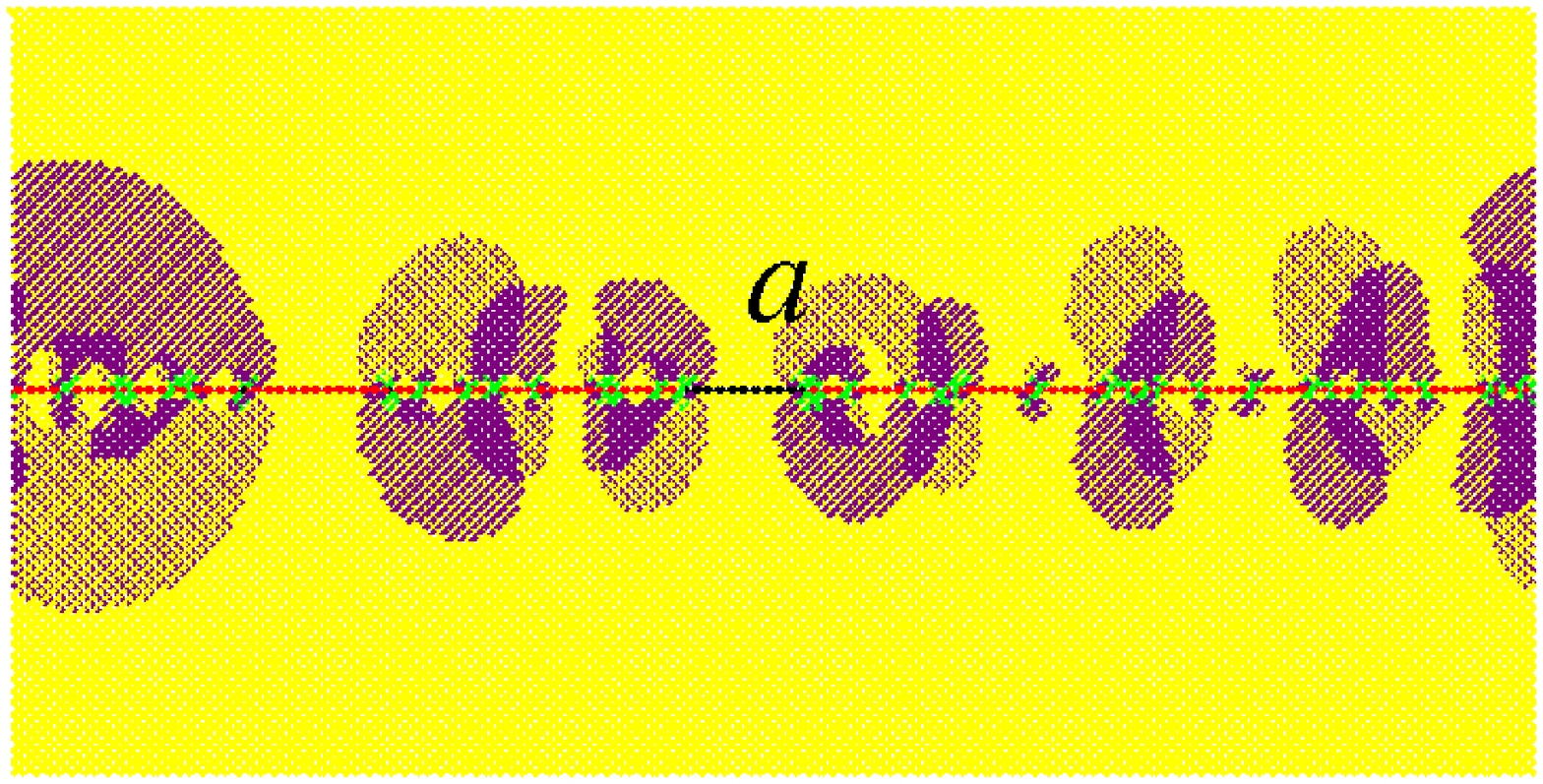}
    \caption{(Color online) Snapshots of the current distribution over
      the lattice for $\gamma=3$ (above) and $\gamma=0$ (below) with
      $b=0.5$ and $L=256$. Initially, the system has a notch at its
      center, which grows horizontally, fracturing the medium in two
      parts. Currents increase as colors change from yellow, over
      violet, to green. The {\it main crack}, the one connected to the
      notch, of length $a$, is colored black, while secondary cracks
      are shown in red.}
    \label{fig:lattice}
  \end{center}
\end{figure}

Only the $L$ resistors in the midline of the system, which includes
the initial notch, are allowed to burn (see Fig.~\ref{fig:lattice}).
Thus, the results reported here are valid as long as the crack tends
to propagate perpendicularly to the direction of the external load.
After a resistor burns, the current distribution is rearranged,
causing an increase in the current around crack ends.  We allow only
one resistor to burn at a time and this burning process continues
until a macroscopic fracture crosses the system. Due to the presence
of disorder, the {\it main crack}, starting from the notch, can grow
either by one resistor at a time or by coalescence with secondary
cracks initiated ahead of the notch. In the latter case, it advances
several resistors instantaneously, stopping at the next resistor which
has not reached its fatigue threshold. This procedure of following a
single one-dimensional crack has been used experimentally to study
growing cracks in thermally activated stressed systems
\cite{Santucci2004}. As Fig.~\ref{fig:lattice} shows, increasing the
value of $\gamma$ greatly reduces the occurrence of secondary cracks.

In the special case of $\gamma=0$, the damage-accumulation rate is
equal for all resistors, and independent on the current amplitude, and
the breaking sequence is determined exclusively by the random
distribution of fatigue thresholds. This particular limit of $\gamma$
is especially important as it enhances the role of disorder in fatigue
failure. The process is then similar to one-dimensional percolation,
and analytical equations can be developed to confirm our results from
numerical simulations.

For $\gamma=0$, one needs only to distinguish between the homogeneous
($b=0$) and finite disorder ($b\neq 0$) cases, since systems with
different values of $b>0$ can be mapped on each other by rescaling
time. Thus, we consider a one-dimensional chain of resistors whose
fatigue thresholds are chosen according to a uniform probability
distribution, $P\left(F^{\mathrm{th}}\right)$, in the interval
$[0,1]$. We define the time unit by imposing that the accumulated
damage at each resistor at time $t$ is $f\left(t\right)=t$, which is
equivalent to measuring time in units of the rupture time. Thus, the
probability that a given resistor is already burnt at time $t$ is
simply $t$. Given that the main crack has length $a$, having started
from a pair of adjacent burnt resistors at time $t=0$, the probability
that it advances $\Delta a$ resistors between times $t$ and $t+dt$,
after having waited a time between $\Delta t$ and $\Delta t+d\left(
  \Delta t\right)$ since it last advanced, is $P\left(t,\Delta
  t,\Delta a|a\right)\, dt\, d\left(\Delta t\right)$, with
\begin{eqnarray}
  P\left(t,\Delta t,\Delta a|a\right) & = & \left(a-2\right)\left(a-1\right)a\left(t-\Delta t\right)^{a-3}\left(1-t\right)
  t^{\Delta a-1}\nonumber\\
  & \times & \left[\left(1-t\right)\left(1-\delta_{\Delta a,L-a-1}\right)+\delta_{\Delta a,L-a-1}\right].
\end{eqnarray}

In order to arrive at the above conditional probability, let us assume
that the crack advances to the right, and notice that $\left(t-\Delta
  t\right)^{a-3}\left(1-t\right)d(\Delta t)$ is the probability that
$a-3$ resistors burned before time $t-\Delta t$, one resistor burned
between $t-\Delta t$ and $t-\Delta t+d(\Delta t)$ (while two resistors
form the initial notch), and the resistor at the left end of the crack
is still intact at time $t$. The factor $t^{\Delta a-1}dt$ corresponds
to the probability that the crack advances $\Delta a$ resistors
between times $t$ and $t+dt$, while the terms in square brackets
differentiate a crack that advances less than $L-a-1$ resistors,
stopping at yet another intact resistor, from a crack that goes around
the system (due to periodic boundary conditions). Of course the
process is symmetric for a crack advancing to the left, and the factor
$\left(a-2\right)\left(a-1\right)a$ ensures normalization of the
conditional probability.

\begin{figure}[t]
  \begin{center}
    \includegraphics*[width=0.93\columnwidth]{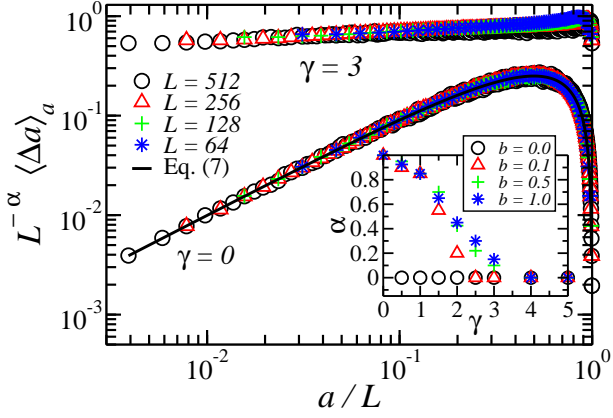}
    \caption{(Color online) Main plot: Rescaled average jump of the
      main crack as a function of its length $a$, for various system
      sizes, two values of the damage accumulation exponent,
      $\gamma=0$ and $\gamma=3$, and disorder strength $b=0.5$. The
      results for $\gamma=0$ match the analytical prediction (solid
      line) from Eq.~(\ref{eq:deltaa}).  The jump of a crack depends
      on its length as a power-law, $\left\langle\Delta
        a\right\rangle_a \propto a^{\alpha}$. The error bars are
      smaller than the symbols. The {\it inset} shows the exponent
      $\alpha$ versus $\gamma$ for different degrees of disorder. In
      the ordered case, $b=0$, the exponent $\alpha$ vanishes for any
      value of $\gamma$ since the crack advances without coalescence.}
    \label{fig:deltaa}
  \end{center}
\end{figure}

The marginal probabilities $P(\Delta a|a)$ and $P(\Delta t|a)$, which
correspond to the distributions of crack jumps and waiting times
between successive jumps, for a given crack length $a$, can be readily
obtained from $P\left(t,\Delta t,\Delta a|a\right)$ by integrating
over the remaining variables, and are given by
\begin{eqnarray}
  P\left(\Delta a|a\right) & = &
  \frac{2\left(a-1\right)a\left(1-\delta_{\Delta a,L-a-1}\right)}
  {\left(a+\Delta a-2\right)\left(a+\Delta a-1\right)\left(a+\Delta a\right)}
  \nonumber \\
  & + &
  \frac{\left(a-1\right)a}{\left(L-3\right)\left(L-2\right)}\delta_{\Delta a,L-a-1}
\end{eqnarray}
and
\begin{equation}
  P\left(\Delta t|a\right)=a\left(1-\Delta t\right)^{a-1}.
\end{equation}
From these, we can calculate, as functions of the crack length $a$,
the average crack jump and the average waiting time between
consecutive jumps,
\begin{equation}
  \langle\Delta a\rangle_a 
  = \sum_{\Delta a=1}^{L-a-1}\Delta a P\left(\Delta a|a\right)
  = a\left(1-\frac{a-1}{L-2}\right)
  \label{eq:deltaa}
\end{equation}
and
\begin{equation}
  \langle\Delta t\rangle_a = \int_{0}^{1}d(\Delta t)\,\Delta t\,P(\Delta t|a)
  = \frac{1}{a+1}.
  \label{eq:deltat}
\end{equation}
Notice that $\langle\Delta a\rangle_a$ grows linearly with $a$ for
$1\ll a\ll L$, but decreases rapidly as $a$ approaches $L$, since
$\Delta a$ cannot be larger than $L-a-1$. On the other hand,
$\langle\Delta t\rangle_a$ decreases as $1/a$ for $a\gg 1$.

\begin{figure}[t]
  \begin{center}
    \includegraphics[width=0.93\columnwidth]{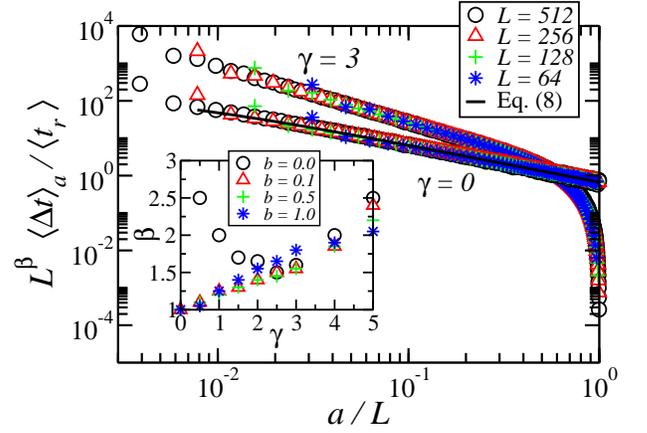}
    \caption{(Color online) Averaged waiting time between successive
      breaks of the main crack, normalized by the averaged rupture
      time, $\left\langle t_r\right\rangle$, as a function of the
      length $a$, collapsed for different system sizes, for $\gamma=0$
      and $\gamma=3$, with $b=0.5$. The results of $\gamma=0$ match
      the analytical prediction (solid line) from
      Eq.~(\ref{eq:deltat}). The waiting-time of a crack depends on
      its length as a power-law, $\left\langle\Delta
        t\right\rangle_a/\left\langle t_r\right\rangle \propto
      a^{-\beta}$. The error bars are smaller than the symbols. The
      {\it inset} shows the exponent $\beta$ versus $\gamma$ for
      different degrees of disorder.}
    \label{fig:deltat}
  \end{center}
\end{figure}

In the presence of disorder, there are various possible definitions
for the crack velocity. For instance, we can determine the average
time dependence of the crack length, $\left\langle a(t)\right\rangle$,
and calculate its derivative. Alternatively, for a given crack length,
we can estimate crack velocity in terms of averages as
\[
\left\langle\frac{\Delta a}{\Delta t}\right\rangle_a,\quad
\frac{\left\langle\Delta a\right\rangle_a}{\left\langle\Delta
    t\right\rangle_a},\quad \mbox{or}\quad\left\langle\frac{\Delta
    t}{\Delta a}\right\rangle^{-1}_a.
\]
However, and not surprisingly, it is possible to show that, for
$\gamma =0$, all these definitions yield crack velocities which scale
with the crack length $a$, for $1\ll a\ll L$, as
\begin{equation}
 v \sim a^2,
\end{equation}
leading, as defined in Eq. (\ref{eq:parislaw}), to a Paris exponent
$m=4$. In our simulations, we estimate the crack velocity, $da/dt$, by
a procedure designed to minimize statistical fluctuations. Precisely,
for a given disorder realization, we estimate the time at which the
crack had a given length. (For disorder realizations in which the
crack happened to be temporarily trapped at that particular length, we
chose the time at which the crack advanced from that length.) Then, we
calculate the average time at each crack length over all disorder
realizations. We take averages over up to 50\,000 realizations,
depending on the set of parameters, $\gamma$, $b$, and $L$. Finally,
we numerically differentiate the corresponding curve with respect to
time, to estimate the crack velocity.

\begin{figure}[t]
  \begin{center}
    \includegraphics[width=0.93\columnwidth]{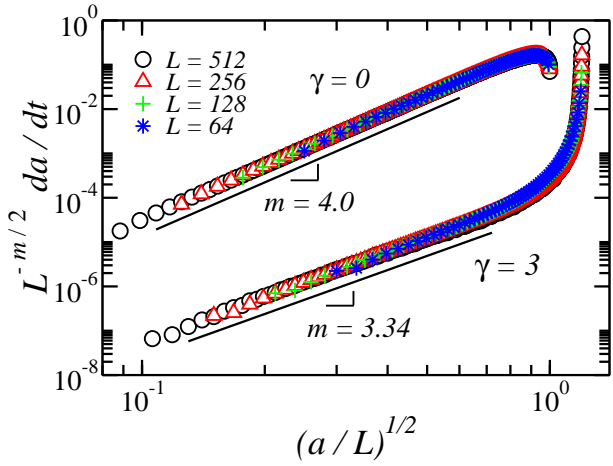}
    \caption{(Color online) Rescaled crack growth rate of the main
      crack as a function of rescaled length $a/L$, showing the Paris
      exponent, $m$, for $\gamma=0$ and $\gamma=3$, with $b=0.5$. For
      better visualization, curves for $\gamma=3$ are shifted to the
      right by $20\%$. }
    \label{fig:paris}
  \end{center}
\end{figure}

We now discuss the results obtained by simulating the fuse-network
model with $\gamma\geq 0$, and compare the case of $\gamma=0$ to the
above analytical expressions. The jumps and waiting times of the main
crack, as functions of the length $a$, are analyzed via finite-size
scaling in Figs.~\ref{fig:deltaa} and \ref{fig:deltat}, respectively.
The results show that both $\left\langle\Delta a\right\rangle_a$ and
$\left\langle\Delta t\right\rangle_a/\left\langle t_r\right\rangle$
scale as power laws of the crack length $a$, $\left\langle\Delta
  a\right\rangle_a \propto a^{\alpha}$ and $\left\langle\Delta
  t\right\rangle_a/\left\langle t_r\right\rangle \propto a^{-\beta}$,
where $\left\langle\Delta t\right\rangle_a$ is normalized by the
average rupture time $\left\langle t_r\right\rangle$. The behaviors of
the exponents $\alpha$ and $\beta$ as functions of the
stress-amplification exponent $\gamma$, for different disorder
strengths $b$, are shown in the insets of Figs.~\ref{fig:deltaa} and
\ref{fig:deltat}, respectively. For $\gamma=0$ and $b>0$, both
$\alpha$ and $\beta$ are unity, in agreement with the analytical
results. For the ordered case ($b=0$), $\alpha$ vanishes for all
values of $\gamma$, since the main crack always advances two resistors
at a time, and no secondary cracks appear. In this limit, $\beta$ is
one half of the Paris exponent $m$.

The growth rate of the main crack is depicted in Fig.~\ref{fig:paris},
which shows the rescaled crack velocity for different system sizes,
disorder strength $b=0.5$, and $\gamma=0$ and $3$. The values of the
Paris exponent are chosen so as to yield the best data collapse of the
curves corresponding to different system sizes for a given value of
$\gamma$. Usual materials present, after the Paris regime, a faster
growth rate preceding the final rupture. However, due to boundary
effects, we observe, for any finite disorder strength ($b>0$) and for
$\gamma$ close to zero, a slight decrease in the growth rate as the
system approaches catastrophic failure.

\begin{figure}[t]
  \begin{center}
    \includegraphics[width=0.9\columnwidth]{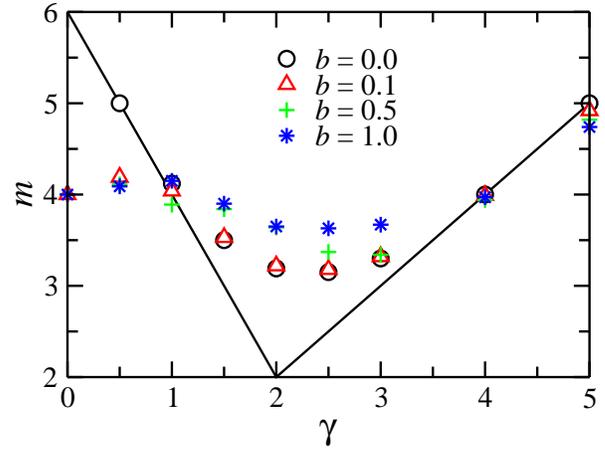}
    \caption{(Color online) The Paris exponent $m$ versus the
      stress-amplification exponent $\gamma$ for different disorder
      strengths $b$. The solid lines represent the results found for
      the homogeneous case, $m=6-2\gamma$, for $\gamma<2$, and
      $m=\gamma$, for $\gamma\geq2$ \cite{Vieira2008}. For the
      one-dimensional case with $\gamma=0$, $m=4$ for any finite value
      of $b$.}
    \label{fig:mxgamma}
  \end{center}
\end{figure}

The dependence of the macroscopic Paris exponent $m$ on the
microscopic fatigue exponent $\gamma$, for different degrees of
disorder, is shown in Fig.~\ref{fig:mxgamma}, together with the results
found for the homogeneous case, $m=6-2\gamma$, for $\gamma<2$, and
$m=\gamma$, for $\gamma\geq2$ \cite{Vieira2008}. Numerical estimates
extracted from fuse-model simulations for the homogeneous case are
also shown, highlighting the existence of logarithmic corrections to
finite-size scaling, which plague the results for values of $\gamma$
around the critical value $\gamma_c=2$.  In the presence of disorder
($b>0$), $m$ converges to $4$ as $\gamma$ approaches zero, as
predicted by the analytical results, while the behavior is similar to
the ordered case for $\gamma>2$, except for very strong disorder
($b\simeq 1$).  Interestingly, any finite disorder leads to a decrease
in the Paris exponent, and thus to a potential increase in the fatigue
lifetime, for $\gamma\lesssim 1$. On the other hand, in agreement with
a stability analysis of the effects of disorder on the homogeneous
system \cite{Vieira2008}, the system is essentially insensitive to
small disorder for $\gamma>2$.

In conclusion, we have investigated both analytically and by
fuse-model simulations the behavior of fatigue failure in
heterogeneous materials, characterized by a damage-accumulation law
governed by a stress amplification exponent $\gamma$. We have
determined how the empirical, macroscopic Paris exponent depends on
the microscopic parameters of fatigue and disorder, represented by
random local fatigue thresholds. For small values of $\gamma$,
corresponding to materials in which damage accumulation depends only
weakly on the local stress amplitude, the Paris exponent approaches
$m=4$ (obtained exactly for $\gamma=0$) if any disorder is present. On
the other hand, fatigue failure in materials for which $\gamma>2$
continues to be characterized by a Paris exponent $m=\gamma$ as long
as disorder is not sufficiently strong.  A considerable challenge for
future research lies in obtaining the relation between atomistic
processes and the coarse-grained description of damage accumulation by
means of a stress-amplification exponent $\gamma$.

\begin{acknowledgments}
  This work was supported by the Brazilian Agencies CNPq, CAPES,
  FINEP, FUNCAP, FAPESP, NAP-FCx, the FUNCAP/CNPq Pronex grant, and
  the National Institute of Science and Technology for Complex
  Systems.
\end{acknowledgments}

\end{document}